\documentclass{article}
\usepackage{graphicx,xcolor,amsmath,amssymb,amsthm,hyperref} 
\usepackage{comment}
\usepackage{circuitikz}

\usepackage{natbib}
\usepackage[margin=3.8cm]{geometry} 

\theoremstyle{definition}

\newtheorem{definition}{Definition}
\newtheorem{ex}{Example}

\newcommand{\tu}[1]{\mathbf{#1}}

\newcommand{\PL}{\mathsf{LTLF}}
\newcommand{\Pg}{\Phi_{\Sigma}}
\newcommand{\Rexp}{R_e}

\newcommand{\vtot}{\mathcal{V}}
\newcommand{\vseg}{\mathbf{a}}

\newcommand{\at}{p}
\newcommand{\fOne}{\phi}
\newcommand{\fTwo}{\psi}
\newcommand{\fThree}{\gamma}

\newcommand{\At}{\mathfrak{P}}

\newcommand{\wbox}{\square}
\newcommand{\bstar}{\triangledown}
\newcommand{\bbox}{\blacksquare}
\newcommand{\wnext}{\triangleright}

\newcommand{\sat}{\circ}

\newcommand{\APL}{\mathsf{CTLF}}
\newcommand{\Rf}{R_f}

\usepackage{minted}
\setminted{breaklines=true}

\title{Counting Worlds Branching Time Semantics for post-hoc Bias Mitigation in generative AI}
\author{Alessandro G. Buda, Giuseppe Primiero, \\ Leonardo Ceragioli, Melissa Antonelli \vspace{1em}\\
\texttt{alessandro.buda@iusppavia.it}\\
\texttt{giuseppe.primiero@unimi.it}\\
\texttt{leonardo.ceragioli@unimi.it}\\
\texttt{melissa.antonelli@uni-tuebingen.de}}
\date{}

\begin{document}

\maketitle

\section*{Abstract}

Generative AI systems are known to amplify biases present in their training data. While several inference-time mitigation strategies have been proposed, they remain largely empirical and lack formal guarantees. In this paper we introduce CTLF, a branching-time logic designed to reason about bias in series of generative AI outputs. CTLF adopts a counting worlds semantics where each world represents a possible output at a given step in the generation process and introduces modal operators that allow us to verify whether the current output series respects an intended probability distribution over a protected attribute, to predict the likelihood of remaining within acceptable bounds as new outputs are generated, and to determine how many outputs are needed to remove in order to restore fairness. We illustrate the framework on a toy example of biased image generation, showing how CTLF formulas can express concrete fairness properties at different points in the output series.

\bigskip

\textit{Keywords}: Computational Tree Logic, Counting World Semantics, Generative AI, Bias Mitigation.

\section{Introduction}

The presence of bias in Generative AI is a widely recognized phenomenon, stemming primarily from the use of unbalanced training datasets. Furthermore, such original bias reflecting real world disparities is amplified by  several factors related to the different components of AI systems, a phenomenon known as bias amplification \cite{bap,bac}. In the case of image generation, one of these factors is the poor representation of certain groups of individuals within the training datasets \cite{Hirota2022}.  Recent surveys provide comprehensive taxonomies of bias sources, evaluation methodologies, and bias mitigation strategies in LLMs \cite{blodgett2024survey,guo2024bias,springer2024survey}. These works categorize mitigation approaches into data-level, model-level, and post-processing techniques, highlighting the growing importance of inference-time control, as retraining is costly and difficult to scale to larger models.

Inference-only and post-hoc mitigation methods aim to reduce bias without retraining foundation models. A number of proposals have been offered in this direction: removing biased activations during inference to mitigate harmful outputs \cite{unibias2024}; token-level filtering mechanisms to enforce fairness constraints during generation \cite{cheng2025biasfilter}; vector ensemble methods to reduce biased generations at inference time \cite{paes2025dso,siddique2025steering}. Simulation-based analyses further evaluate the effectiveness of such inference-time interventions under controlled bias scenarios \cite{simulation2025}.  Guardrail architectures provide an external safety layer around LLMs modularly combining content classification, policy enforcement, and fallback mechanisms to ensure trustworthy deployment \cite{bridging2024}. Broader reviews discuss layered safety frameworks and architectural constraints designed to intercept unsafe or biased outputs before user delivery \cite{ayyamperumal2024currentstatellmrisks,syed2025guardrails}. 
Empirical studies also examine bias in LLM-based bias detection systems themselves, revealing disparities between model judgments and human perception \cite{lin2025investigating}. 

From the point of view of formal analysis and verification, it is useful to check whether, given a prompt with an expected distribution of a protected attribute at any point in the output production, is possible to determine bias amplification beyond a given acceptable threshold. When it is already known that the output will become unacceptable, continuing using the model may become useless or harmful. This requires predicting the evolution of the output series as early as possible. Following such verification process, one may want to start a post-hoc bias mitigation process which will work on the current output to remove portions of it in order to bring the distribution back within admissible limits. This requires to reason temporally on series of outputs and their possible extensions. The linear temporal logic $\PL$ offered in \cite{LFS} allows to infer frequencies of events on future states of the output. In the present paper, we present $\APL$ an alternative counting world CTL-style strategy to establish whether at any current state of a series of outputs the expected final series will lay behind the acceptable threshold for unbiased outputs, and if so how to intervene to mitigate the situation as  early as possible.

\section{A Toy Example}
\begin{figure}[!ht]
    \centering
    \includegraphics[width=0.9\textwidth]{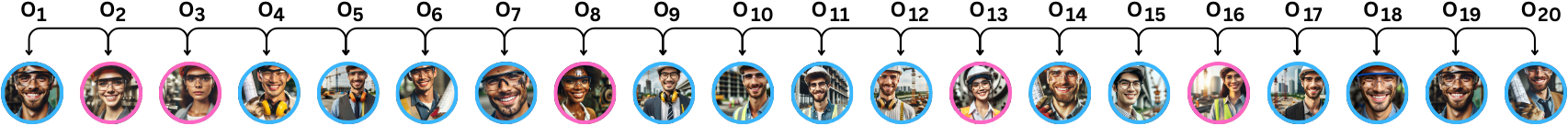}
    \caption{Original Training Set, which will inform the Output's Distribution}
    \label{fig:D}
\end{figure}

Consider a dataset $D$ of $20$ images with an unfair gender distribution (e.g., 75\% M/25\% F), see Figure \ref{fig:D}. Leaving aside any amplification, we assume that when $D$ is used to train a model $M$ the output series $O=\{o_{1}, \dots, o_{n}\}$ 
on the prompt \textit{``Give me a picture of an engineer"} repeated $n$ times will reflect the same gender distribution of $D$. Knowing the current and expected distribution of the output and its cardinality, the objective is to mitigate $O$ at inference stage to obtain an output subset $O'\subset O$ 
with cardinality as close to $n$ as possible with a desired gender distribution (e.g., $50\%M/50\% F$). Hence, we formulate the following questions:
\begin{figure}
    \centering
\includegraphics[width=1\textwidth]{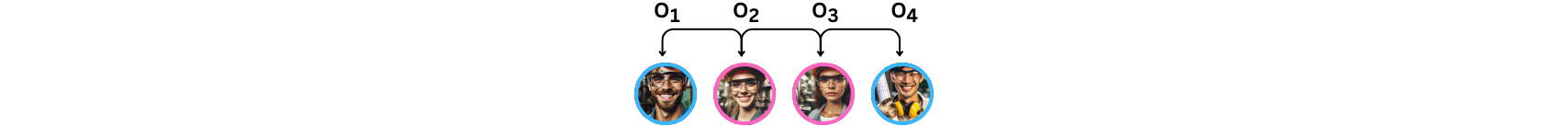}
    \caption{A Subset of the Output at the fourth item in the series.}
    \label{fig:Q1}
\end{figure}
\begin{enumerate}
\itemsep0em
    \item At output $m<n$, are we still within the intended distribution (e.g. a fair one)? See Figure \ref{fig:Q1}
    
    \item If so, what is the probability to be within the intended  distribution at output $m'$ s.t.  $m<m'\leq n$?

     \item[3] If not, how many outputs do we need to remove to restore it? See Figure \ref{fig:Q2}
\end{enumerate}

\begin{figure}[!h]

    \begin{center}
    \includegraphics[width=1\textwidth]{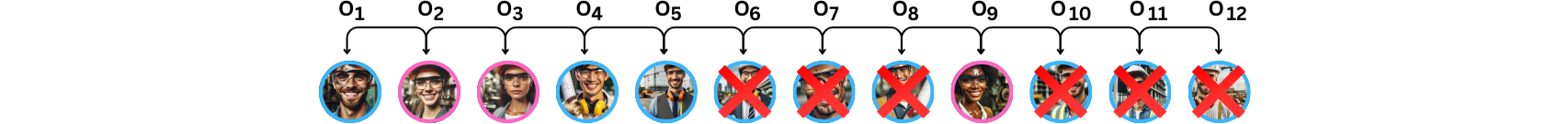}

    \includegraphics[width=1\textwidth]{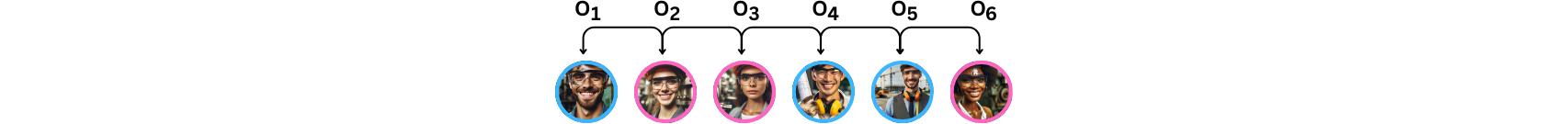}
    \caption{A mitigated Output subset.}
    \label{fig:Q2}
    \end{center}
\end{figure}

In the following example we will refer to odds for the sake of readability, however it can be easily extended to frequencies and/or probabilities. Consider again Figure \ref{fig:D} as training where the initial odds are $15/5 = 3/1$ in favor of males. Given the results up to o$_4$ (see Figure \ref{fig:Q1}), we are still within the intended (say fair) distribution of the gender property. To establish if the results are still within the intended distribution at o$_7$, one requires calculating the odds of the output being male at all intermediate steps:
\begin{enumerate}
    \item At $o_5$ they are $13/3 > 4/1$; 
    \item At $o_6$ they can be $12/3 = 4/1$ if we obtained a male at $o_5$, or $13/2 > 6/1$ if we obtained a female subject;
    \item At $o_7$ they can be $11/3 < 4/1$ if we obtained a male subject both at $o_5$ and $o_6$, or $12/2 = 6/1$ if we obtained a male at $o_5$, a female at $o_6$, and \textit{vice versa}, or $13/1$ if we obtained a female subject both at $o_5$ and $o_6$.
\end{enumerate}

Consider now the case depicted in Figure \ref{fig:Q2}, where twelve outputs have been generated within the same unfair distribution of the training dataset ($9/3 = 3/1$ in favor of males). At $o_5$ the result is $3/2$ in favor of male subjects. Given that we have already reached the maximum number of male subjects within the intended distribution for the set of generated outcomes, we should start mitigating by removing all the following outcomes that will result in male engineers.

Our aim is to provide a semantics to reason logically about such distributions, their extensions and required mitigation when the desired output is no longer achieved.

\section{The Logic $\APL$ and its counting semantics}
The logic $\PL$ is introduced in \cite{LFS} to reason about series of probabilistic events. We present an extension $\APL$ with additional operators and an alternative semantics. In what follows, we will use $\at_1, \at_2, \dots$ for atomic propositions. We assume that such atoms are linguistic representations of our probabilistic events; accordingly, as events are mutually exclusive and collectively exhaustive, the occurrence of an event (i.e. an atom in the language) implies the negation of any other event within the same sample space (i.e. a negation of an atom is equivalent to the disjunction of all other atoms in the same sample space). For example, for $p$="male", $\lnot p\leftrightarrow q$="female", or $\lnot p\leftrightarrow q\vee r$="female or trans". $\fOne, \fTwo, \fThree, \dots$ denote possibly molecular propositions. We use $\At$ to denote the set of propositional variables.

\begin{definition}[Language of $\APL$]
    Formulas of $\APL$ are defined by the grammar below:
    $$
    \fOne ::= \at \mid \neg \fOne \mid \fOne \wedge \fOne \mid \fOne \vee \fOne \mid \wbox_q \fOne \mid \bbox_q \fOne \mid \sat_q \fOne \mid \wnext_{q} \fOne 
    $$
    $$
    \fTwo :=  \bstar_{q} \fOne \mid \dagger_{q}\fOne
    $$
    where $\at \in \At$ and $q\in \mathbb{Q}_{[0,1]}$.
\end{definition}

In the following, we rely on standard probability theory terminology.
Intuitively, an event denotes an output on a prompt; a series of events denotes a series of outputs of size $n$; the outcome or value of an event denotes the occurrence of a given (possibly protected) property in an output. A world can be though as the occurrence of an event.

$\fOne$ denotes a state formula. A propositional variable $\at$ is meant to denote linguistically the value of a protected attribute of output $o$ of model $\mathcal{M}$, and it can be closed under negation (useful for binary protected attributes), conjunction and disjunction. We call this an \textit{event}. Given a series of $n$ events, $\wbox_{q} \fOne$ says that at the current state $\fOne$ has happened \emph{at least} $q \times m$ times, where $m\leq n$ is the number of outputs so far. The formula $\bbox_{q} \fOne$ expresses that at the current state $\fOne$ could have happened (possibly in a different series) at least $q \times m$ times, where $m\leq n$ is the number of outputs so far within a well-defined probability distribution $\Pg$. The formula $\sat _{q}\fOne$ expresses that among the possible complete series of outputs, there is one such series in which at some point $\fOne$ has happened at least $q \times n$ times, where $n$ is the total number of outputs happening in that series. $\wnext_{q} \fOne$ expresses that, given a complete series of outputs within a well-defined probability distribution, and a point in that series, the probability that $\fOne$ happens at the next time in that series is at least $q$, namely $q \times (n-m)$ times. The formula $\fTwo$ is a path formula. The formula $\bstar_{q}\fOne$ expresses that given a complete series of events $\fOne$ happens at least $q\times n$ times. 
The formula $\dagger_{q}\fOne$ expresses that starting at a given event of a complete series with outputs in a well-defined probability distribution, the probability that such series is completed with outputs still in that distribution is at least $q$.

\begin{definition}[Model of $\APL$]\label{def:AltPrL}
Let $\Sigma$ be a well-defined probability distribution on an event space. We define a model
$\mathcal{M} = (W, W^1, W^n,\Rexp, \Rf, \vtot)$, where
\begin{itemize}
\itemsep0em
    \item $W=\{w_{1.1}, w_{1.2}, \dots, w_{1.l}, w_{2.1}, w_{2.2}, \dots, w_{2.l \times l}, \dots w_{n.1}, \dots, w_{n.m}\}$ is a finite set of worlds s.t.~$l$ is the number of possible outcomes for each event, $n$ is teh number of events for each series and $m=l^n$ is the number of possible series of events.
    
    \item $W\supset W^1 = \{w_{1.1}, w_{1.2}, \dots, w_{1.l}\}$ is the set of \textit{root worlds}, denoting the set of possible values for the first event of the series. Its  cardinality is $l$;

    \item $W\supset W^n = \{w_{n.1}, w_{n.2}, \dots, w_{n.m}\}$ is the set of \textit{leaf worlds}, \textit{i.e.}, the set of possible outcomes for the last event of the series, with cardinality $m$;
    
    \item $\Rexp$ is an anti-symmetric, beginning, reflexive, and transitive (backward) relation, \textit{i.e.}, a \textit{childhood} relation between different events of the series, such that:
    \begin{itemize}
            \item $\forall w \in W^{1}, w'\Rexp w \iff w' = w$: root events have no parents;

        \item for each pair $w_{i.x}, w_{j,y} \in W$ if  $w_{i.x}\Rexp w_{j.y}$ then $i\geq j$: non-root events are children of previous ones 
        %
        \item for each $w_{i.j}\in W$ with $i >1$, $\exists ! w_{i-1.k} : w_{i.j} \Rexp w_{i-1.k}$ with $k = \lceil j/l \rceil$, and, as seen, $l$ is the number of possible outcomes for each event: each non-root event $w_{i.j}$ at time $i>1$ can have only one \textit{direct} parent world at time $i-1$, whose secondary index is given by the ceiling function of $j/l$. 
        Therefore, given a series of $n$ events with $l$ possible outcomes for each event, there exists only one path 
        from each of the possible $l^n$ outcomes for the last event through all events in the series to the root world.
    \end{itemize}

    \item $\Rf$ is an anti-symmetric, beginning, reflexive, and transitive (forward) relation, \textit{i.e.}, a \textit{parenthood} relation between different events of the series, such that:
     \begin{itemize}
             \item $\forall w\in W^n, w'\Rf w \iff w' = w$: leaf events have no children;

        \item for each pair $w_{i.x},w_{j,y} \in W$, if $w_{i.x}\Rf w_{j.y}$ then $i\leq j$: previous events are seen as parents of following ones;
        %
        \item for each $w_{i.j} \in W$ with $i < n$ , there exists a set $\{ w_{i+1.h+1}, w_{i+1.h+2}, \dots,$ $ w_{i+1.h+l}\}$ with $h=[(j-1)\times l]$, and $l$ is the number of possible outcomes for each event, \textit{i.e.}, each non-leaf world $w_{i.j}$ representing one possible outcome for the event at time $i<n$ has $l$ \textit{direct} children at time $i+1$, with secondary indexes $[(j-1)\times l]+1, [(j-1)\times l]+2, \dots, [(j-1)\times l]+l$. 
        Therefore, given a series of $n$ events, with $l$ possible outcomes for each event, and given a non leaf world $w_{i.j}$, representing the $j$th possible outcome for the $i$th event in the series, there are $l^{n-i}$ possible path from $w_{i.j}$ to the leaf worlds, passing across all the intermediate events.
    \end{itemize}
    
    \item 
    $\vtot: W \mapsto \mathcal{P}(\At)$ is a valuation function assigning to each world in $W$ a  subset of $\At$ holding in such world. 


\end{itemize}
\end{definition}

We denote with $W^i = \{w_{i.1}, w_{i.2}, \dots , w_{i.l}\}$ the set of possible outcomes for the $i$th event of the series.
We now introduce the notion of path.
\begin{definition}[Path]\label{def:path}
Given $W$ and $\Rf \subseteq W \times W$, we define a \textit{path} between events $i$ and $j$ of the series (with $i < j$) as:
\[
    \pi_{[i.x,j.z]}=\{ \langle w_{i.x}, w_{i+1.y}, \dots, w_{j.z}\rangle : w_{k.v} \in W, (w_{k.v}, w_{k+1.w}) \in \Rf, \forall k \in [i,j-1] \}
\]
\end{definition}

\noindent
An equivalent definition can be provided using $\Rexp$ instead of $\Rf$. A \textit{complete path} $\pi^*$ is a path from the first to the last event of the series, including all the intermediate steps:
$
\pi^*_{[x,z]}=\pi_{[1.x,n.z]} 
$
\begin{definition}[Set of paths]\label{def:setofpaths}
Given $W$ and $\Rf \subseteq W \times W$, and a world $w_{i.j}$, we define the \textit{set of paths} originating in $w_{i.j}$ as:
$
\Pi^{[i.j]} = \{ \langle w_{i.j}, \dots, w_{n.k}\rangle,$ $ \dots, $ $\langle w_{i.j}, $ $ \dots, $ $ w_{n.m}\rangle\}. \
$ 
The set 
$
\Pi^* = \{ \langle w_{1.1}, \dots, w_{n.k}\rangle, \dots ,\langle w_{1.l}, \dots, w_{n.m}\rangle\}
$ 
is the \textit{set of complete paths}. 
\end{definition}


\begin{definition}[$\Pg$]\label{def:AltPg}
Given a probability distribution $\Sigma$ assigning to any $p \in \At$ the frequency $\mu (p)$, the function $\Pg: \Pi^* \mapsto \mathcal{P}(\Pi^*)$  associates the set of complete paths in $W$ it returns any of its member $\pi^*_{[x,z]}$ such that 
\[
\frac{\vert \{ w_{i.j} \in \pi^*_{[x,z]} : w_{i.j} \in \vtot(p)\}\vert}{\vert \pi^*_{[x,z]} \vert}= \mu (p)
\]
\end{definition}

\noindent
In other words, every path in the co-domain of $\Pg$ shows for each $p \in \At$ a set of $\mu (p) \times \vert \pi^* \vert$ worlds in which $p$ is true.
%
%
We denote with $\Pi^{[i.j]}_{\Sigma}$
the set of paths originating in $w_{i.j}$ returned by $\Pg$, 
%
and with $\Pi^*_{\Sigma}$
the set of complete paths returned by  $\Pg$.
We denote with $v:\pi_{[i.x,j.z]}\mapsto \mathcal{P}(\mathfrak{P})$ any function for the series of events corresponding to path  $\pi_{[i.x,j.z]}$, identifying one among the possible $m=l^n$ series of events, or possibly complete paths.

\begin{definition}[Reachable Worlds]

Given a model $\mathcal{M}$, for any world $w_{i.j} \in W$ we denote:
\begin{itemize}
    \item $\Rexp^{w_{i.j}} = \{w' \in W: w_{i.j} \Rexp w'\}$, the set of all worlds accessible from $w_{i.j}$ according to the transitive closure on $\Rexp$.
   
    \item $\Rf^{w_{i.j}} = \{w' \in W: w_{i.j} \Rf w'\}$, the set of all worlds accessible from $w_{i.j}$ according to the transitive closure on $\Rf$.

\end{itemize}
\end{definition}

\begin{definition}[Semantics for state  and path formulas]\label{def:prop_semantics}
For any $\APL$-formula $\fOne$, a model $\mathcal{M}$, a world $w\in W$,  we define the relation $\mathcal{M}, w \vDash$ inductively as follows:
\begin{flalign*}
&\mathcal{M}, w_{i.j} \vDash \at \iff w_{i.j} \in \tu \vtot(\at)\\
&\mathcal{M}, w_{i.j} \vDash \neg \fOne \iff M, w_{i.j} \not\vDash \fOne\\
& \mathcal{M}, w_{i.j} \vDash \fOne_{1} \wedge \fOne_{2} \iff M, w_{i.j} \vDash \fOne_{1}\ \text{ and }\ M, w_{i.j} \vDash \fOne_{2} \\
& \mathcal{M}, w_{i.j} \vDash \fOne_{1} \vee \fOne_{2} \iff M, w_{i.j} \vDash \fOne_{1}\ \text{ or }\ M, w_{i.j} \vDash \fOne_{2} \\
& \mathcal{M}, w_{i.j} \vDash \wbox_{q} \fOne  \iff  \frac{|\Rexp^{w_{i.j}} (\fOne)|}{|\Rexp^{w_{i.j}}|} \ge q. \\
& \mathcal{M}, w_{i.j} \vDash \bbox_{q} \fOne \iff\! 
\frac{|\Rexp^{w_{i.k}}(\fOne)|}{|\Rexp^{w_{i.k}}|} \ge q\ 
\text{ for some } \pi^{*}_{\Sigma} \text{ with }\! w_{i.k\leqslant \geqslant j} \in \pi^*_{\Sigma} \text{ and } w_{i.k\leqslant \geqslant j} \in W^{i}\!.\\
& \mathcal{M}, w_{i.j} \vDash \sat_{q} \fOne \iff 
\frac{|\Rexp^{w_{i.j}} (\fOne)|}{|\pi^*|} \geq q\ \text{ for some } \pi^{*}. \\
& \mathcal{M}, w_{i.j} \vDash \wnext_{q} \fOne \iff 
\frac{\mid \Rf^{w_{i+1.k}} (\fOne)\mid}{\mid \Rf^{w_{i+1.k}}\mid} \geq q\ \text{ for some }  \pi^*_{\Sigma} \text{ such that } w_{i.j}, w_{i+1, k} \in \pi^*_{\Sigma}, 
\end{flalign*}
where $\Rexp^{w_{i.j}} (\fOne) = \{ w' \in \Rexp^{w_{i.j}} : \mathcal{M}, w_{i.j} \vDash\phi \}$ and $\Rf^{w_{i.j}} (\fOne) = \{ w' \in \Rf^{w_{i.j}} :  \mathcal{M}, w_{i.j} \vDash \phi \}$
%
%
For any $\APL$-formula $\fOne$, a model $\mathcal{M}$, a (possibly complete) path $\pi_{[i.x,j.z]}$,  we define the relation $\mathcal{M}, \pi_{[i.x,j.z]} \vDash$ inductively as follows:
\begin{flalign*}
& \mathcal{M}, \pi^{*}_{[x,z]} \vDash \bstar_{q} \fOne \iff \frac{\mid \Rf^{w_{1.x}} (\fOne)\mid}{\mid\pi^*_{[x,z]} \mid} \ge q\\
& \mathcal{M},\pi_{[i.x,j.z]} \vDash \dagger_{q}\fOne\iff\mathcal{M}, \pi^{*}_{[x,z]}\supseteq \pi_{[i.x,j.z]} \vDash \bstar_{q'} \fOne \text{ with } \pi^{*}_{[x,z]} \in \Pi^*_{\Sigma}
\text{ and } \frac{\mid \Pi^{[j.z]}_{\Sigma} \mid}{\mid \Pi^{[j.z]}\mid}\geq q. 
\end{flalign*}
\end{definition}

\section{Formalising our Example}

\begin{ex}
    Consider the model in Fig. \ref{fig:AltRexp} representing $n=6$ generations of images on the prompt \textit{``Give me a picture of an engineer"}, with $l=2$ possible outcomes ($M$ and $F$, for the sake of simplicity) for each generation, and $m=2^6=64$ possible series of events. Connections between outcomes represent both the experimental ($\Rexp$), and the forward-looking ($\Rf$) relations. Paths with valuations in $\Pg$, \textit{i.e.}, paths in $\Pi^*_{\Sigma}$, are marked with dashed lines.
    $$W=\{w_{1.1}, w_{1.2}, w_{2.1},\dots, w_{2.4}, w_{3.1},\dots, w_{3.8}, w_{4.1},\dots, w_{4.16}, w_{5.1},\dots, w_{5.32},w_{6.1},\dots, w_{6.64}\}$$
    
    Worlds $w_{1.1}$ and $w_{1.2}$ are the root worlds. For each \textit{non-root} world there exists only one direct parent world belonging to the set of possible outcomes for the previous step in the series (\textit{e.g.}, world $w_{3.5}$ has only one direct parent world, $w_{(3-1).\lceil 5/2 \rceil}=w_{2.3}$), and there is only one complete path for each world connecting this to the corresponding root world (\textit{e.g.}, from $w_{3.5}$ to $w_{2.3}$, and then to $w_{1.2}$).
    The subset of $W$ $\{w_{6.1}, \dots, w_{6.64}\}$ is the set of leaf worlds. For each non leaf world there are $l=2$ direct children worlds, representing the possible outcomes for the next event of the series (e.g. $w_{2.3}$ has two direct children, $w_{2+1.[(3-1)\times 2]+1} = w_{3.5}$, and $w_{2+1.[(3-1)\times 2]+2} = w_{3.6}$), and there are $2^{6-2}=16$ possible path from $w_{2.3}$ to the corresponding leafs (from $w_{2.3}$ to $w_{6.33}, w_{6.34}, w_{6.35}, \dots w_{6.48}$.

    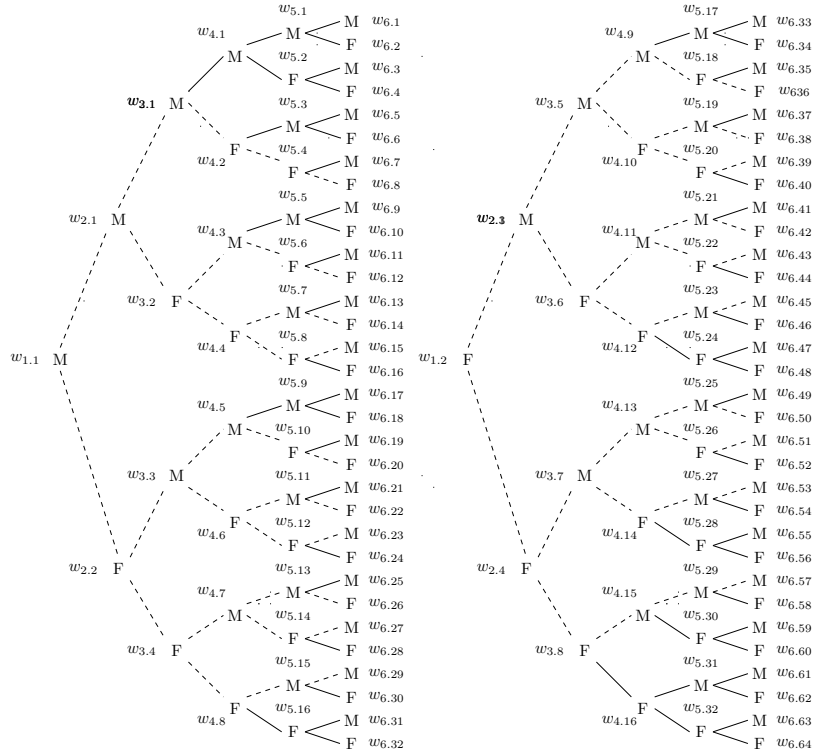
\begin{figure}[!ht]\label{fig:model}
    \centering
    \resizebox{.8\textwidth}{!}{%
        \begin{circuitikz}
    \tikzstyle{every node}=[font=\normalsize]
    \node [font=\normalsize] at (7.5,18.5) {M};
    \node [font=\normalsize] at (7.5,17) {F};
    \node [font=\normalsize] at (7.5,18) {F};
    \node [font=\normalsize] at (7.5,17.5) {M};
    \node [font=\normalsize] at (7.5,16.5) {M};
    \node [font=\normalsize] at (7.5,15) {F};
    \node [font=\normalsize] at (7.5,16) {F};
    \node [font=\normalsize] at (7.5,15.5) {M};
    \node [font=\normalsize] at (7.5,14.5) {M};
    \node [font=\normalsize] at (7.5,13) {F};
    \node [font=\normalsize] at (7.5,14) {F};
    \node [font=\normalsize] at (7.5,13.5) {M};
    \node [font=\normalsize] at (7.5,12.5) {M};
    \node [font=\normalsize] at (7.5,11) {F};
    \node [font=\normalsize] at (7.5,12) {F};
    \node [font=\normalsize] at (7.5,11.5) {M};
    \node [font=\normalsize] at (7.5,10.5) {M};
    \node [font=\normalsize] at (7.5,9) {F};
    \node [font=\normalsize] at (7.5,10) {F};
    \node [font=\normalsize] at (7.5,9.5) {M};
    \node [font=\normalsize] at (7.5,8.5) {M};
    \node [font=\normalsize] at (7.5,7) {F};
    \node [font=\normalsize] at (7.5,8) {F};
    \node [font=\normalsize] at (7.5,7.5) {M};
    \node [font=\normalsize] at (7.5,6.5) {M};
    \node [font=\normalsize] at (7.5,5) {F};
    \node [font=\normalsize] at (7.5,6) {F};
    \node [font=\normalsize] at (7.5,5.5) {M};
    \node [font=\normalsize] at (7.5,4.5) {M};
    \node [font=\normalsize] at (7.5,3) {F};
    \node [font=\normalsize] at (7.5,4) {F};
    \node [font=\normalsize] at (7.5,3.5) {M};
    \node [font=\normalsize] at (1.25,11.25) {M};
    \node [font=\normalsize] at (2.5,14.25) {M};
    \node [font=\normalsize] at (2.5,6.75) {F};
    \node [font=\normalsize] at (6.25,18.25) {M};
    \node [font=\normalsize] at (6.25,15.25) {F};
    \node [font=\normalsize] at (6.25,17.25) {F};
    \node [font=\normalsize] at (6.25,16.25) {M};
    \node [font=\normalsize] at (6.25,14.25) {M};
    \node [font=\normalsize] at (6.25,11.25) {F};
    \node [font=\normalsize] at (6.25,13.25) {F};
    \node [font=\normalsize] at (6.25,12.25) {M};
    \node [font=\normalsize] at (6.25,10.25) {M};
    \node [font=\normalsize] at (6.25,7.25) {F};
    \node [font=\normalsize] at (6.25,9.25) {F};
    \node [font=\normalsize] at (6.25,8.25) {M};
    \node [font=\normalsize] at (6.25,6.25) {M};
    \node [font=\normalsize] at (6.25,3.25) {F};
    \node [font=\normalsize] at (6.25,5.25) {F};
    \node [font=\normalsize] at (6.25,4.25) {M};
    \node [font=\normalsize] at (5,17.75) {M};
    \node [font=\normalsize] at (5,11.75) {F};
    \node [font=\normalsize] at (5,15.75) {F};
    \node [font=\normalsize] at (5,13.75) {M};
    \node [font=\normalsize] at (5,9.75) {M};
    \node [font=\normalsize] at (5,3.75) {F};
    \node [font=\normalsize] at (5,7.75) {F};
    \node [font=\normalsize] at (5,5.75) {M};
    \node [font=\normalsize] at (3.75,16.75) {M};
    \node [font=\normalsize] at (3.75,12.5) {F};
    \node [font=\normalsize] at (3.75,8.75) {M};
    \node [font=\normalsize] at (3.75,5) {F};
    \draw [ dashed, short] (1.75,12.5) -- (1.75,12.5);
    \draw [ dashed, short] (1.25,11.5) -- (2.25,14);
    \draw [ dashed, short] (2.5,14.5) -- (3.5,16.5);
    \draw [ dashed, short] (4,16.75) -- (4.75,16);
    \draw [ dashed, short] (4.25,16.25) -- (4.25,16.25);
    \draw [ dashed, short] (5.25,15.75) -- (6,15.5);
    \draw [ dashed, short] (6.5,15.25) -- (7.25,15);
    \draw [ dashed, short] (1.25,11) -- (2.5,7);
    \draw [ dashed, short] (2.75,14) -- (3.5,12.75);
    \draw [ dashed, short] (2.75,7) -- (3.5,8.5);
    \draw [ dashed, short] (2.75,6.5) -- (3.5,5.25);
    \draw [ dashed, short] (4,9) -- (4.75,9.75);
    \draw [ dashed, short] (5.25,9.75) -- (6,9.5);
    \draw [ dashed, short] (6.5,9.25) -- (7.25,9);
    \draw [ dashed, short] (4,12.5) -- (4.75,12);
    \draw [ dashed, short] (5.25,11.75) -- (6,11.25);
    \draw [ dashed, short] (6.5,11.25) -- (7.25,11.5);
    \draw [ dashed, short] (6,11.5) -- (6,11.5);
    \draw [ dashed, short] (5.5,11.5) -- (5.5,11.5);
    \draw [ dashed, short] (4,12.75) -- (4.75,13.5);
    \draw [ dashed, short] (5.25,13.75) -- (6,13.5);
    \draw [ dashed, short] (6.5,13.25) -- (7.25,13);
    \draw [ dashed, short] (6,13.25) -- (6,13.25);
    \draw [dashed] (4,17) -- (4,17);
    \draw [dashed] (6.75,18.25) -- (6.75,18.25);
    \draw [dashed] (9,17.75) -- (9,17.75);
    \draw [short] (4,17) -- (4.75,17.75);
    \draw [short] (5.25,18) -- (6,18.25);
    \draw [short] (6.5,18.25) -- (7.25,18.5);
    \draw [short] (6.75,18) -- (6.75,18);
    \draw [short] (5.25,17.75) -- (6,17.25);
    \draw [short] (6.5,18.25) -- (7.25,18);
    \draw [short] (6.5,17.25) -- (7.25,17.5);
    \draw [short] (6.5,17.25) -- (7.25,17);
    \draw [short] (6.5,16.25) -- (7.25,16.5);
    \draw [short] (6.5,16.25) -- (7.25,16);
    \draw [short] (5.25,16) -- (6,16.25);
    \draw [short] (7.25,16) -- (7.25,16);
    \draw [short] (7,16) -- (7,16);
    \draw [short] (6.5,15.25) -- (7.25,15.5);
    \draw [short] (9.25,15.75) -- (9.25,15.75);
    \draw [short] (5.25,14) -- (6,14.25);
    \draw [short] (6.5,14.25) -- (7.25,14.5);
    \draw [short] (6.5,14.25) -- (7.25,14);
    \draw [short] (6.5,13.25) -- (7.25,13.5);
    \draw [short] (6.5,11.25) -- (7.25,11);
    \draw [ dashed, short] (5.25,12) -- (6,12.25);
    \draw [ dashed, short] (6.5,12.25) -- (7.25,12);
    \draw [short] (6.5,12.25) -- (7.25,12.5);
    \draw [short] (5.25,9.75) -- (5.25,9.75);
    \draw [short] (5.25,10) -- (6,10.25);
    \draw [short] (6.5,10.25) -- (7.25,10.5);
    \draw [short] (6.5,10.25) -- (7.25,10);
    \draw [short] (6.5,9.25) -- (7.25,9.5);
    \draw [ dashed, short] (4,8.5) -- (4.75,8);
    \draw [ dashed, short] (5.25,8) -- (6,8.25);
    \draw [ dashed, short] (6.5,8.25) -- (7.25,8);
    \draw [ dashed, short] (5.25,7.75) -- (6,7.25);
    \draw [ dashed, short] (6.5,7.25) -- (7.25,7.5);
    \draw [ dashed, short] (4,5.25) -- (4.75,5.75);
    \draw [ dashed, short] (5.25,6) -- (6,6.25);
    \draw [ dashed, short] (6.5,6.25) -- (7.25,6);
    \draw [ dashed, short] (5.5,6) -- (5.5,6);
    \draw [ dashed, short] (5.5,6) -- (5.5,6);
    \draw [ dashed, short] (5.25,5.75) -- (6,5.25);
    \draw [ dashed, short] (6.5,5.25) -- (7.25,5.5);
    \draw [ dashed, short] (4,4.75) -- (4.75,4);
    \draw [ dashed, short] (5.25,4) -- (6,4.25);
    \draw [ dashed, short] (6.5,4.25) -- (7.25,4.5);
    \draw [ dashed, short] (9.25,8.5) -- (9.25,8.5);
    \draw [short] (9,8.75) -- (9,8.75);
    \draw [short] (6.5,8.25) -- (7.25,8.5);
    \draw [short] (6.5,7.25) -- (7.25,7);
    \draw [short] (6.5,6.25) -- (7.25,6.5);
    \draw [short] (6.5,5.25) -- (7.25,5);
    \draw [short] (6.5,4.25) -- (7.25,4);
    \draw [short] (6.5,3.25) -- (7.25,3.5);
    \draw [short] (6.5,3.25) -- (7.25,3);
    \draw [short] (5.25,3.75) -- (6,3.25);
    \draw [short] (5.75,6.25) -- (5.75,6.25);
    \draw [short] (5.75,6) -- (5.75,6);
    \node [font=\normalsize] at (16.25,18.5) {M};
    \node [font=\normalsize] at (16.25,17) {F};
    \node [font=\normalsize] at (16.25,18) {F};
    \node [font=\normalsize] at (16.25,17.5) {M};
    \node [font=\normalsize] at (16.25,16.5) {M};
    \node [font=\normalsize] at (16.25,15) {F};
    \node [font=\normalsize] at (16.25,16) {F};
    \node [font=\normalsize] at (16.25,15.5) {M};
    \node [font=\normalsize] at (16.25,14.5) {M};
    \node [font=\normalsize] at (16.25,13) {F};
    \node [font=\normalsize] at (16.25,14) {F};
    \node [font=\normalsize] at (16.25,13.5) {M};
    \node [font=\normalsize] at (16.25,12.5) {M};
    \node [font=\normalsize] at (16.25,11) {F};
    \node [font=\normalsize] at (16.25,12) {F};
    \node [font=\normalsize] at (16.25,11.5) {M};
    \node [font=\normalsize] at (16.25,10.5) {M};
    \node [font=\normalsize] at (16.25,9) {F};
    \node [font=\normalsize] at (16.25,10) {F};
    \node [font=\normalsize] at (16.25,9.5) {M};
    \node [font=\normalsize] at (16.25,8.5) {M};
    \node [font=\normalsize] at (16.25,7) {F};
    \node [font=\normalsize] at (16.25,8) {F};
    \node [font=\normalsize] at (16.25,7.5) {M};
    \node [font=\normalsize] at (16.25,6.5) {M};
    \node [font=\normalsize] at (16.25,5) {F};
    \node [font=\normalsize] at (16.25,6) {F};
    \node [font=\normalsize] at (16.25,5.5) {M};
    \node [font=\normalsize] at (16.25,4.5) {M};
    \node [font=\normalsize] at (16.25,3) {F};
    \node [font=\normalsize] at (16.25,4) {F};
    \node [font=\normalsize] at (16.25,3.5) {M};
    \node [font=\normalsize] at (10,11.25) {F};
    \node [font=\normalsize] at (11.25,14.25) {M};
    \node [font=\normalsize] at (11.25,6.75) {F};
    \node [font=\normalsize] at (15,18.25) {M};
    \node [font=\normalsize] at (15,15.25) {F};
    \node [font=\normalsize] at (15,17.25) {F};
    \node [font=\normalsize] at (15,16.25) {M};
    \node [font=\normalsize] at (15,14.25) {M};
    \node [font=\normalsize] at (15,11.25) {F};
    \node [font=\normalsize] at (15,13.25) {F};
    \node [font=\normalsize] at (15,12.25) {M};
    \node [font=\normalsize] at (15,10.25) {M};
    \node [font=\normalsize] at (15,7.25) {F};
    \node [font=\normalsize] at (15,9.25) {F};
    \node [font=\normalsize] at (15,8.25) {M};
    \node [font=\normalsize] at (15,6.25) {M};
    \node [font=\normalsize] at (15,3.25) {F};
    \node [font=\normalsize] at (15,5.25) {F};
    \node [font=\normalsize] at (15,4.25) {M};
    \node [font=\normalsize] at (13.75,17.75) {M};
    \node [font=\normalsize] at (13.75,11.75) {F};
    \node [font=\normalsize] at (13.75,15.75) {F};
    \node [font=\normalsize] at (13.75,13.75) {M};
    \node [font=\normalsize] at (13.75,9.75) {M};
    \node [font=\normalsize] at (13.75,3.75) {F};
    \node [font=\normalsize] at (13.75,7.75) {F};
    \node [font=\normalsize] at (13.75,5.75) {M};
    \node [font=\normalsize] at (12.5,16.75) {M};
    \node [font=\normalsize] at (12.5,12.5) {F};
    \node [font=\normalsize] at (12.5,8.75) {M};
    \node [font=\normalsize] at (12.5,5) {F};
    \draw [dashed, short] (10.5,12.5) -- (10.5,12.5);
    \draw [dashed, short] (10,11.5) -- (11,14);
    \draw [dashed, short] (11.25,14.5) -- (12.25,16.5);
    \draw [dashed, short] (12.75,16.75) -- (13.5,16);
    \draw [dashed, short] (13,16.25) -- (13,16.25);
    \draw [dashed, short] (14,15.75) -- (14.75,15.5);
    \draw [short] (15.25,15.25) -- (16,15);
    \draw [dashed, short] (10,11) -- (11.25,7);
    \draw [dashed, short] (11.5,14) -- (12.25,12.75);
    \draw [dashed, short] (11.5,7) -- (12.25,8.5);
    \draw [dashed, short] (11.5,6.5) -- (12.25,5.25);
    \draw [dashed, short] (12.75,9) -- (13.5,9.75);
    \draw [dashed, short] (14,9.75) -- (14.75,9.5);
    \draw [short] (15.25,9.25) -- (16,9);
    \draw [dashed, short] (12.75,12.5) -- (13.5,12);
    \draw [short] (14,11.75) -- (14.75,11.25);
    \draw [short] (15.25,11.25) -- (16,11.5);
    \draw [dashed, short] (14.75,11.5) -- (14.75,11.5);
    \draw [dashed, short] (14.25,11.5) -- (14.25,11.5);
    \draw [dashed, short] (12.75,12.75) -- (13.5,13.5);
    \draw [dashed, short] (14,13.75) -- (14.75,13.5);
    \draw [short] (15.25,13.25) -- (16,13);
    \draw [dashed, short] (14.75,13.25) -- (14.75,13.25);
    \draw [dashed] (12.75,17) -- (12.75,17);
    \draw [dashed] (15.5,18.25) -- (15.5,18.25);
    \draw [dashed] (17.75,17.75) -- (17.75,17.75);
    \draw [dashed, short] (12.75,17) -- (13.5,17.75);
    \draw [short] (14,18) -- (14.75,18.25);
    \draw [short] (15.25,18.25) -- (16,18.5);
    \draw [short] (15.5,18) -- (15.5,18);
    \draw [dashed, short] (14,17.75) -- (14.75,17.25);
    \draw [short] (15.25,18.25) -- (16,18);
    \draw [short] (15.25,17.25) -- (16,17.5);
    \draw [dashed, short] (15.25,17.25) -- (16,17);
    \draw [short] (15.25,16.25) -- (16,16.5);
    \draw [dashed, short] (15.25,16.25) -- (16,16);
    \draw [dashed, short] (14,16) -- (14.75,16.25);
    \draw [short] (16,16) -- (16,16);
    \draw [short] (15.75,16) -- (15.75,16);
    \draw [dashed, short] (15.25,15.25) -- (16,15.5);
    \draw [short] (18,15.75) -- (18,15.75);
    \draw [dashed, short] (14,14) -- (14.75,14.25);
    \draw [short] (15.25,14.25) -- (16,14.5);
    \draw [dashed, short] (15.25,14.25) -- (16,14);
    \draw [dashed, short] (15.25,13.25) -- (16,13.5);
    \draw [short] (15.25,11.25) -- (16,11);
    \draw [dashed, short] (14,12) -- (14.75,12.25);
    \draw [short] (15.25,12.25) -- (16,12);
    \draw [dashed, short] (15.25,12.25) -- (16,12.5);
    \draw [short] (14,9.75) -- (14,9.75);
    \draw [dashed, short] (14,10) -- (14.75,10.25);
    \draw [short] (15.25,10.25) -- (16,10.5);
    \draw [dashed, short] (15.25,10.25) -- (16,10);
    \draw [dashed, short] (15.25,9.25) -- (16,9.5);
    \draw [dashed, short] (12.75,8.5) -- (13.5,8);
    \draw [dashed, short] (14,8) -- (14.75,8.25);
    \draw [short] (15.25,8.25) -- (16,8);
    \draw [short] (14,7.75) -- (14.75,7.25);
    \draw [short] (15.25,7.25) -- (16,7.5);
    \draw [dashed, short] (12.75,5.25) -- (13.5,5.75);
    \draw [dashed, short] (14,6) -- (14.75,6.25);
    \draw [short] (15.25,6.25) -- (16,6);
    \draw [dashed, short] (14.25,6) -- (14.25,6);
    \draw [dashed, short] (14.25,6) -- (14.25,6);
    \draw [short] (14,5.75) -- (14.75,5.25);
    \draw [short] (15.25,5.25) -- (16,5.5);
    \draw [short] (12.75,4.75) -- (13.5,4);
    \draw [short] (14,4) -- (14.75,4.25);
    \draw [short] (15.25,4.25) -- (16,4.5);
    \draw [dashed, short] (18,8.5) -- (18,8.5);
    \draw [short] (17.75,8.75) -- (17.75,8.75);
    \draw [dashed, short] (15.25,8.25) -- (16,8.5);
    \draw [short] (15.25,7.25) -- (16,7);
    \draw [dashed, short] (15.25,6.25) -- (16,6.5);
    \draw [short] (15.25,5.25) -- (16,5);
    \draw [short] (15.25,4.25) -- (16,4);
    \draw [short] (15.25,3.25) -- (16,3.5);
    \draw [short] (15.25,3.25) -- (16,3);
    \draw [short] (14,3.75) -- (14.75,3.25);
    \draw [short] (14.5,6.25) -- (14.5,6.25);
    \draw [short] (14.5,6) -- (14.5,6);
    \node [font=\normalsize] at (0.5,11.25) {$w_{1.1}$};
    \node [font=\normalsize] at (9.25,11.25) {$w_{1.2}$};
    \node [font=\normalsize] at (1.75,14.25) {$w_{2.1}$};
    \node [font=\normalsize] at (1.75,6.75) {$w_{2.2}$};
    \node [font=\normalsize] at (10.5,6.75) {$w_{2.4}$};
    \node [font=\normalsize] at (10.5,14.25) {$w_{2.1}$};
    \node [font=\normalsize] at (10.5,14.25) {$w_{2.3}$};
    \node [font=\normalsize] at (3,12.5) {$w_{3.2}$};
    \node [font=\normalsize] at (3,16.75) {$w_{2.1}$};
    \node [font=\normalsize] at (3,16.75) {$w_{3.1}$};
    \node [font=\normalsize] at (3,5) {$w_{3.4}$};
    \node [font=\normalsize] at (3,8.75) {$w_{3.3}$};
    \node [font=\normalsize] at (11.75,8.75) {$w_{3.7}$};
    \node [font=\normalsize] at (11.75,12.5) {$w_{3.6}$};
    \node [font=\normalsize] at (11.75,16.75) {$w_{3.5}$};
    \node [font=\normalsize] at (11.75,5) {$w_{3.8}$};
    \node [font=\normalsize] at (4.5,14) {$w_{4.3}$};
    \node [font=\normalsize] at (4.5,15.5) {$w_{4.2}$};
    \node [font=\normalsize] at (4.5,18.25) {$w_{4.1}$};
    \node [font=\normalsize] at (4.5,11.5) {$w_{4.4}$};
    \node [font=\normalsize] at (4.5,10.25) {$w_{4.5}$};
    \node [font=\normalsize] at (4.5,7.5) {$w_{4.6}$};
    \node [font=\normalsize] at (4.5,6.25) {$w_{4.7}$};
    \node [font=\normalsize] at (4.5,3.5) {$w_{4.8}$};
    \node [font=\normalsize] at (13.25,18.25) {$w_{4.9}$};
    \node [font=\normalsize] at (13.25,15.5) {$w_{4.10}$};
    \node [font=\normalsize] at (13.25,14) {$w_{4.11}$};
    \node [font=\normalsize] at (13.25,11.5) {$w_{4.12}$};
    \node [font=\normalsize] at (13.25,10.25) {$w_{4.13}$};
    \node [font=\normalsize] at (13.25,7.5) {$w_{4.14}$};
    \node [font=\normalsize] at (13.25,6.25) {$w_{4.15}$};
    \node [font=\normalsize] at (13.25,3.5) {$w_{4.16}$};
    \node [font=\normalsize] at (6.25,18.75) {$w_{5.1}$};
    \node [font=\normalsize] at (6.25,17.75) {$w_{5.2}$};
    \node [font=\normalsize] at (6.25,16.75) {$w_{5.3}$};
    \node [font=\normalsize] at (6.25,15.75) {$w_{5.4}$};
    \node [font=\normalsize] at (6.25,14.75) {$w_{5.5}$};
    \node [font=\normalsize] at (6.25,13.75) {$w_{5.6}$};
    \node [font=\normalsize] at (6.25,12.75) {$w_{5.7}$};
    \node [font=\normalsize] at (6.25,11.75) {$w_{5.8}$};
    \node [font=\normalsize] at (6.25,3.75) {$w_{5.16}$};
    \node [font=\normalsize] at (6.25,4.75) {$w_{5.15}$};
    \node [font=\normalsize] at (6.25,5.75) {$w_{5.14}$};
    \node [font=\normalsize] at (6.25,6.75) {$w_{5.13}$};
    \node [font=\normalsize] at (6.25,7.75) {$w_{5.12}$};
    \node [font=\normalsize] at (6.25,8.75) {$w_{5.11}$};
    \node [font=\normalsize] at (6.25,9.75) {$w_{5.10}$};
    \node [font=\normalsize] at (6.25,10.75) {$w_{5.9}$};
    \node [font=\normalsize] at (15,18.75) {$w_{5.17}$};
    \node [font=\normalsize] at (15,17.75) {$w_{5.18}$};
    \node [font=\normalsize] at (15,16.75) {$w_{5.19}$};
    \node [font=\normalsize] at (15,15.75) {$w_{5.20}$};
    \node [font=\normalsize] at (15,14.75) {$w_{5.21}$};
    \node [font=\normalsize] at (15,13.75) {$w_{5.22}$};
    \node [font=\normalsize] at (15,12.75) {$w_{5.23}$};
    \node [font=\normalsize] at (15,11.75) {$w_{5.24}$};
    \node [font=\normalsize] at (15,10.75) {$w_{5.25}$};
    \node [font=\normalsize] at (15,9.75) {$w_{5.26}$};
    \node [font=\normalsize] at (15,8.75) {$w_{5.27}$};
    \node [font=\normalsize] at (15,7.75) {$w_{5.28}$};
    \node [font=\normalsize] at (15,6.75) {$w_{5.29}$};
    \node [font=\normalsize] at (15,5.75) {$w_{5.30}$};
    \node [font=\normalsize] at (15,4.75) {$w_{5.31}$};
    \node [font=\normalsize] at (15,3.75) {$w_{5.32}$};
    \node [font=\normalsize] at (8.25,18.5) {$w_{6.1}$};
    \node [font=\normalsize] at (8.25,18) {$w_{6.2}$};
    \node [font=\normalsize] at (8.25,17.5) {$w_{6.3}$};
    \node [font=\normalsize] at (8.25,17) {$w_{6.4}$};
    \node [font=\normalsize] at (8.25,16.5) {$w_{6.5}$};
    \node [font=\normalsize] at (8.25,16) {$w_{6.6}$};
    \node [font=\normalsize] at (8.25,15.5) {$w_{6.7}$};
    \node [font=\normalsize] at (8.25,15) {$w_{6.8}$};
    \node [font=\normalsize] at (8.25,14) {$w_{6.10}$};
    \node [font=\normalsize] at (8.25,13.5) {$w_{6.11}$};
    \node [font=\normalsize] at (8.25,13) {$w_{6.12}$};
    \node [font=\normalsize] at (8.25,12.5) {$w_{6.13}$};
    \node [font=\normalsize] at (8.25,12) {$w_{6.14}$};
    \node [font=\normalsize] at (8.25,11.5) {$w_{6.15}$};
    \node [font=\normalsize] at (8.25,11) {$w_{6.16}$};
    \node [font=\normalsize] at (8.25,14.5) {$w_{6.9}$};
    \node [font=\normalsize] at (8.25,10.5) {$w_{6.17}$};
    \node [font=\normalsize] at (8.25,10) {$w_{6.18}$};
    \node [font=\normalsize] at (8.25,9.5) {$w_{6.19}$};
    \node [font=\normalsize] at (8.25,9) {$w_{6.20}$};
    \node [font=\normalsize] at (8.25,8.5) {$w_{6.21}$};
    \node [font=\normalsize] at (8.25,8) {$w_{6.22}$};
    \node [font=\normalsize] at (8.25,7.5) {$w_{6.23}$};
    \node [font=\normalsize] at (8.25,7) {$w_{6.24}$};
    \node [font=\normalsize] at (8.25,6) {$w_{6.26}$};
    \node [font=\normalsize] at (8.25,5.5) {$w_{6.27}$};
    \node [font=\normalsize] at (8.25,5) {$w_{6.28}$};
    \node [font=\normalsize] at (8.25,4.5) {$w_{6.29}$};
    \node [font=\normalsize] at (8.25,4) {$w_{6.30}$};
    \node [font=\normalsize] at (8.25,3.5) {$w_{6.31}$};
    \node [font=\normalsize] at (8.25,3) {$w_{6.32}$};
    \node [font=\normalsize] at (8.25,6.5) {$w_{6.25}$};
    \node [font=\normalsize] at (17,18.5) {$w_{6.33}$};
    \node [font=\normalsize] at (17,18) {$w_{6.34}$};
    \node [font=\normalsize] at (17,17.5) {$w_{6.35}$};
    \node [font=\normalsize] at (17,17) {$w_{636}$};
    \node [font=\normalsize] at (17,16.5) {$w_{6.37}$};
    \node [font=\normalsize] at (17,16) {$w_{6.38}$};
    \node [font=\normalsize] at (17,15.5) {$w_{6.39}$};
    \node [font=\normalsize] at (17,15) {$w_{6.40}$};
    \node [font=\normalsize] at (17,14) {$w_{6.42}$};
    \node [font=\normalsize] at (17,13.5) {$w_{6.43}$};
    \node [font=\normalsize] at (17,13) {$w_{6.44}$};
    \node [font=\normalsize] at (17,12.5) {$w_{6.45}$};
    \node [font=\normalsize] at (17,12) {$w_{6.46}$};
    \node [font=\normalsize] at (17,11.5) {$w_{6.47}$};
    \node [font=\normalsize] at (17,11) {$w_{6.48}$};
    \node [font=\normalsize] at (17,14.5) {$w_{6.41}$};
    \node [font=\normalsize] at (17,10.5) {$w_{6.49}$};
    \node [font=\normalsize] at (17,10) {$w_{6.50}$};
    \node [font=\normalsize] at (17,9.5) {$w_{6.51}$};
    \node [font=\normalsize] at (17,9) {$w_{6.52}$};
    \node [font=\normalsize] at (17,8.5) {$w_{6.53}$};
    \node [font=\normalsize] at (17,8) {$w_{6.54}$};
    \node [font=\normalsize] at (17,7.5) {$w_{6.55}$};
    \node [font=\normalsize] at (17,7) {$w_{6.56}$};
    \node [font=\normalsize] at (17,6) {$w_{6.58}$};
    \node [font=\normalsize] at (17,5.5) {$w_{6.59}$};
    \node [font=\normalsize] at (17,5) {$w_{6.60}$};
    \node [font=\normalsize] at (17,4.5) {$w_{6.61}$};
    \node [font=\normalsize] at (17,4) {$w_{6.62}$};
    \node [font=\normalsize] at (17,3.5) {$w_{6.63}$};
    \node [font=\normalsize] at (17,3) {$w_{6.64}$};
    \node [font=\normalsize] at (17,6.5) {$w_{6.57}$};
    \end{circuitikz}
    }%
     \caption{A model representing 6 generations of images on the prompt \textit{``Give me a picture of an engineer"}. Connections represent both the experimental ($\Rexp$), and the forward-looking ($\Rf$) relations. Paths in $\Pi^*_{\Sigma}$ are marked with dashed lines.}
    \label{fig:AltRexp}
    \end{figure}
\end{ex}

\begin{ex}
Consider again the model in Figure \ref{fig:AltRexp}. The following hold:
\begin{itemize}
\itemsep0em
    \item $w_{1.1} \vDash M$ because $w_{1.1}$ is in $\tu \vseg(M)$, \textit{i.e.}, $M$ is true in $w_{1.1}$
    
    \item $w_{1.1} \vDash \neg F$ because $w_{1.1} \not\vDash F$, \textit{i.e.}, $F$ is false in $w_{1.1}$
    
    \item $w_{3.1} \vDash \wbox_{1} M$ because $|\Rexp^{w_{3.1}} (M)|\ / \ |\Rexp^{w_{3.1}}| = 1 $
    
    \item $w_{3.2} \vDash \wbox_{2/3} M$ because $|\Rexp^{w_{3.1}} (M)|\ / \ |\Rexp^{w_{3.1}}| = 2/3$
    
    \item $w_{3.1} \vDash \bbox_{1}F$ and $w_{3.2} \vDash \bbox_{1}F$ because there exists  $\pi^*_{\Sigma} = \{\langle w_{1.2},\dots,w_{6.57} \rangle \}$ and there exists $w_{3.8}$ belonging to both $\pi^*_{\Sigma}$ and $ W^3$ such that $|\Rexp^{w_{3.8}} (F)| \ / \ |\Rexp^{w_{3.8}}| = 1$
    
    \item $w_{3.1} \vDash \sat_{3/6} M \land \sat_{0/6} F$ because there exists $\pi^*$ \textit{, e.g.}, $\{\langle w_{[1.1]} \dots w_{[6.1]} \rangle\}$, such that $w_{3.1}$ is in $\pi^*$, $|\Rexp^{w_{3.1}} (M)| \ / \ |\pi^*| = 3/6$, and $|\Rexp^{w_{3.1}} (F)| \ / \ |\pi^*| = 0 $
    
    \item $w_{3.1} \vDash \wnext_{1} F$ because there exists $\pi^*_{\Sigma} = \{\langle w_{1.1},\dots,w_{6.8} \rangle \}$ such that $w_{3.1}, w_{4.2}$ are in $\pi^*_{\Sigma}$ and $|\Rf^{w_{4.2}} (F)| \ / \ |\Rf^{w_{4.2}}| = 1 $
    
    \item $w_{3.2} \vDash \wnext_{2/3} F$ because there exists $\pi^*_{\Sigma} = \{\langle w_{1.1},\dots,w_{6.12} \rangle \}$  such that $w_{3.2}, w_{4.3}$ are in $\pi^*_{\Sigma}$ and $|\Rf^{w_{4.3}} (F)| \ / \ |\Rf^{w_{4.3}}| = 2/3$

    \item $\pi^{*}_{[2,63]} \vDash \bstar_{1/6} M$ because starting from the root world $w_{1.2}$ we can reach, by means of $R_f$ across the path $\pi^{*}_{[2,63]}$, only one world over six in which $M$ holds, formally $|\Rf^{w_{1.2}} (M)| \ / \ |\pi^{*}_{[2,63]}| = 1/6$    
    
    \item $\pi_{[1.1,3.1]} \vDash \dagger_{1/8}M$ because $\pi_{[1.1,3.1]}$ is one of the paths originating in $w_{1.1}$ returned by $\Pg$, and there is only one possible path over eight starting from $w_{3.1}$ returned by $\Pg$.
\end{itemize}

\end{ex}

\begin{ex}
Consider the results depicted in Figure \ref{fig:Q1}. This situation can be represented in the model in Figure \ref{fig:AltRexp} as the path starting from $w_{1.1}$ (thus we can ignore the set of paths $\Pi^{[1.2]}$), passing across $w_{[2.2]}$ and $w_{[3.4]}$, and reaching $w_{[4.7]}$. This sub-series of event is still within a fair distribution, in fact we are in a path in $\Pi^*_{\Sigma}$ (a dashed path in the figure). Using the defined language: 
$
w_{4.7}\vDash \wbox_{1/2}M \wedge \wbox_{1/2}F. \
$
Both $M$ and $F$ have appeared in proportion $\geq 1/2$ among the ancestors, hence the output is balanced so far.

We can further characterise the situation at $w_{4.7}$ in terms of the probability of completing the series within a fair distribution. The path $\pi_{[1.1,4.7]}$ is one of the paths originating from $w_{1.1}$ returned by $\Phi_\Sigma$, and from $w_{4.7}$ there are $l^{n-i} = 2^2 = 4$ possible continuations, of which only 2 yield a $\Sigma$-compatible completion reaching $3M/3F$ at event 6. Since $|\Pi^{[4.7]}_\Sigma|\,/\,|\Pi^{[4.7]}| = 2/4 = 1/2$, we have: $ \pi_{[1.1,4.7]} \vDash \dagger_{1/2} M$. 
This means that, given we are currently on a $\Sigma$-compatible path, the probability of completing the series within the fair distribution is exactly $1/2$. Any single deviation at step~5 or~6 eliminates all remaining $\Sigma$-compatible completions.

The two $\Sigma$-compatible completions can be further characterised using the forward-looking
operator $\rhd$. They are $w_{4.7} \to w_{5.13}(M) \to w_{6.26}(F)$ and
$w_{4.7} \to w_{5.14}(F) \to w_{6.27}(M)$. In both cases
$|R_f^{w_{5,k}}(\phi)|\,/\,|R_f^{w_{5,k}}| = 1/2$, yielding:
$ w_{4.7} \vDash \rhd_{1/2}F \;\wedge\; \rhd_{1/2}M $. Both next outputs preserve a fair completion, but each forces a unique outcome at step~6:
the series is balanced but offers no redundancy.

Consider now the mitigation scenario depicted in Figure~\ref{fig:Q2}, where 12 outputs are generated following the training distribution ($9/3 = 3/1$ in favour of males). Since at most 3 female outputs can appear in the full series, the largest achievable balanced subset has cardinality~6, yielding $|\pi^*| = 6$ for the fair-subset path. At $o_5$ the ratio is $3/2$ in favour of males: as $|R_e^{w_{5,j}}(M)| = 3 = 1/2 \times |\pi^*|$, we have: $w_{5,j} \vDash \circ_{1/2}M$, meaning the male quota for the balanced subset is exhausted. Mitigation is nonetheless possible: there exists a $\Sigma'$-compatible path with target distribution $50\%M/50\%F$ passing through $w_{5,j}$ such that $\pi^*_{\Sigma'} \vDash \nabla_{1/2}F$, confirming that a $3M/3F$ subset is achievable from our current position. The mitigation strategy consists in removing all subsequent male outputs (marked $\times$ in Figure~3), yielding a balanced subset $O' \subset O$ of 6 outputs ($\pi_{[1.1,6.26]}$ in Figure \ref{fig:AltRexp}).
\end{ex}

\section{Conclusions}

Our modalities were originally inspired by the counting semantics from \cite{LEGASTELOIS2017} and its use for a pragmatic of information \cite{Buda2025-BUDALF}, by probabilistic interpreted systems \cite{DBLP:journals/sLogica/ChenPRR16}, by probability logics~\cite{Nilsson,Bacchus,FHM}, as well as by the counting operators introduced in counting logics, which were mostly developed in connection with complexity~\cite{Wagner,APAL24,ADLP} and finite model theory~\cite{Moskowakis,Kontinen}. The strategy of checking againts admissible distributions is at the basis of a proof-theoretic and possible-worlds semantics for trustworthiness evaluation and preservation of non-deterministic computations \cite{DBLP:journals/ijar/KubyshkinaP24,DBLP:journals/logcom/DAsaroGP25,DBLP:journals/ijar/CeragioliP26}

Several directions for future work are worth pursuing. First, the current model treats all transitions within a given probability assignemnt as equally probable. A natural extension would assign explicit weights to the transitions, reflecting prior knowledge about the generation distribution. This would allow a more fine-grained account of the likelihood of individual outputs, and could be integrated into the semantics of the counting operators via a weighted reachability measure, enriching the expressive power of the logic without altering its overall structure.
Second, a more significant extension concerns the case in which the reference probability distribution is not fixed in advance, but is refined at runtime as new outputs are observed. Rather than assuming the Markov chain as a known input, one could start from an initial distribution and update the transition probabilities each time a new output is produced: if a "male" output is generated, the weight on the transition towards output "female" is slightly increased for the subsequent step, progressively steering the generation towards the intended distribution. This setting is closely related to the study of parametric Markov chains \cite{Hutschenreiter_2017}, where transition probabilities are expressed as functions of a set of parameters that can be synthesised or updated to satisfy a given specification \cite{Jansen2022}, and to recent work on the formal verification of Markov processes with learned parameters \cite{maaz2025formalverificationmarkovprocesses}. The resulting framework would be an adaptive model in which the counting worlds structure is progressively identified from observations. Establishing formal guarantees on the convergence of such updates and on the correctness of the verification and mitigation procedures in this dynamic setting constitutes an open problem of independent interest.
Finally, a software implementation of the checking and mitigation procedures described in this paper is under development for the MIRAI Toolbox, which has already been applied to several classification tasks \cite{DBLP:conf/beware/CoragliaDGGPPQ23,DBLP:conf/beware/BudaCGMP24,DBLP:conf/hhai/BudaCGMP25}. The tool will take as input a stream of generative AI outputs, a target distribution and a fairness threshold, and will return either a certificate of compliance or a mitigation plan specifying which outputs to remove. This will allow the framework to be evaluated on real-world systems, including both text-to-image models and large language models. 
\section*{Acknowledgments}
 This research was supported by the Ministero dell’Università e della Ricerca (MUR) through PRIN 2022 Project SMARTEST – Simulation of Probabilistic Systems for the Age of the Digital Twin (20223E8Y4X), the Project “Departments of Excellence 2023-2027” awarded to the Department of Philosophy “Piero Martinetti” of the University of Milan which provided funding for a research visit in 2025 to the last author.

\section*{Declaration on Generative AI}
   
 \noindent
 During the preparation of this work, the authors used DALL-E (OpenAI) for figures 1, 2 and 3 in order to: Generate images. After using these tool(s)/service(s), the author(s) reviewed and edited the content as needed and take(s) full responsibility for the publication’s content. 

\bibliographystyle{abbrv}
\bibliography{references}

\end{document}